\documentclass[%
 reprint,
 amsmath,amssymb,
 aps,prb,superscriptaddress
]{revtex4-2}

\usepackage{graphicx}
\usepackage{dcolumn}
\usepackage{bm}
\usepackage{hyperref}
\usepackage{physics}
\usepackage{bbm}
\usepackage{xcolor}

\usepackage{nicefrac}
\usepackage{braket}

\begin{document}

\preprint{APS/123-QED}

\title{How Landau caterpillars turn into Hofstadter butterflies by tuning the periodic potential strength}

\author{Ivo A. Gabrovski}
 \email{ivo.gabrovski@unige.ch}
 \affiliation{Department of Quantum Matter Physics, University of Geneva, 24 quai Ernest-Ansermet, 1211 Geneva, Switzerland}
\author{Louk Rademaker}%
 \affiliation{Department of Quantum Matter Physics, University of Geneva, 24 quai Ernest-Ansermet, 1211 Geneva, Switzerland}%
 \affiliation{Institute-Lorentz for Theoretical Physics, Leiden University, PO Box 9506, 2300 Leiden, The Netherlands}

\date{\today}

\begin{abstract}
It is well-known that the spectrum of two-dimensional electrons in a perpendicular magnetic field is given by discrete flat Landau levels. 
By contrast, electrons in a two-dimensional tight-binding model give rise to a fractal Hofstadter butterfly spectrum.
In this paper, we connect these two opposite limits by showing how a butterfly spectrum emerges from broadened Landau `caterpillars', by  continuously increasing the periodic potential strength.
We identify a series of topological transitions that isolate a lowest trivial band, a necessary condition for the butterfly to emerge. The resulting butterfly is topologically distinct from the Hofstadter butterfly at fluxes $\phi>1$, due to anomalous behavior of diagonal hopping. 
Moreover, the hopping parameters of an effective tight-binding model, obtained by a Wannierization procedure at large magnetic field, are highly dependent on the flux, revealing that Wannier orbitals themselves change under the applied magnetic field.
Our methods and results are relevant for artificial materials, such as moiré systems, where a full quantum of flux per lattice unit cell is experimentally accessible. On a theoretical level, having Wannierization methods for each specific flux allows more accurate many-body calculations in large magnetic fields.
\end{abstract}

\maketitle


\section{\label{sec:intro}Introduction}

The behavior of electrons in crystalline solids under an applied magnetic field is a topic of great interest since the early days of quantum mechanics, starting with the discovery of Shubnikov-de Haas and de Haas-van Alphen effect \cite{dehaas1930dependence}. The observed oscillations of resistivity and magnetization can be understood in terms of a quantization of the spectrum into discrete Landau levels \cite{landau1930diamagnetism}. In two-dimensional materials, the periodic crystal lattice further reorganizes the electronic spectrum into subbands whose fractal structure resembles a butterfly \cite{hofstadter_energy_1976}. However, direct observations of the Hofstadter butterfly have been out of reach. For this, one needs to achieve on the order of one flux quantum per unit cell, which in a conventional atomic-scale crystal requires a magnetic field of $\mathcal{O}(10^4)$ T. This is roughly an order of magnitude larger than the record obtained using destructive electromagnetic flux compression \cite{nakamura2018record} and more than two orders of magnitude larger than the largest direct-current magnetic fields \cite{hahn2019tesla}. 

This experimental obstacle is overcome with the recent advances in moiré materials \cite{Kennes.2021}, which are two-dimensional engineered heterostructures whose lattice constants are on the order of several nanometers. In this case, one flux quantum per unit cell is experimentally realizable with fields in the range of 20 - 30 T. Indeed, transport measurements in moiré materials such as twisted bilayer graphene have revealed butterfly physics through characteristic features such as quantum Hall gaps and Landau fans \cite{dean_hofstadters_2013,yu_hierarchy_2014,saito_hofstadter_2021}. Related spectra have also been pursued in systems with artificial gauge fields, including cold atoms \cite{jaksch_creation_2003}, microwave structures \cite{kuhl_microwave_1998}, and polaritonic lattices \cite{banerjee_realization_2018}. 

These experimental developments stimulate us to revisit the theory of electrons in the presence of both a magnetic field and a periodic potential. The goal of this paper is to fill in the gap between two historically distinct limits: discrete Landau levels on the one hand, which neglects the crystal lattice completely; and the Hofstadter butterfly, which starts from a tight-binding hopping model neglecting Landau levels. 

Starting from the Landau level perspective, inspired by experiments \cite{shoenberg_evidence_1969}, it was shown that a weak cosine periodic potential in two dimensions broadens and splits each Landau level into magnetic subbands \cite{rauh_bloch_1974,rauh_degeneracy_1974,rauh_broadening_1975,thouless_quantized_1982}. In the opposite Hofstadter limit, a sufficiently deep periodic potential produces Bloch bands that can be described by a tight-binding model, with the magnetic field introduced solely through gauge-dependent hopping phases \cite{peierls1933zur,harper_single_1955,hofstadter_energy_1976}. These two limits are related but not identical. The tight-binding butterfly is periodic in the flux through a unit cell, whereas the weak-potential splitting of an isolated Landau level displays the corresponding structure as a function of the \textit{inverse} flux \cite{thouless_quantized_1982,macdonald_landau-level_1983}! So while the equations for the Hofstadter butterfly and the Landau level broadening look very similar, the flux versus inverse flux behavior already reveal these two limits cannot be simply connected. 

When both the magnetic field and the periodic potential are strong, the spectrum is controlled by competition between magnetic and lattice orbital length scales, and neither limit can be treated via crude approximations. We show that once the magnetic length becomes comparable to the spatial extent of a Wannier orbital, the field will affect the orbital shape and overlap rather than contributing only a hopping phase like is done in the Hofstadter model. Conversely, when the periodic potential is strong, the Bloch spectrum is clearly different from that of an unbounded, broadened Landau level, resulting in a series of topological transitions in the spectrum. Note that in moir\'e systems, and twisted bilayer graphene in particular, approaches starting from these two opposite limits can consequently produce fundamentally different spectra \cite{bistritzer_moire_2011,zhang_landau_2019,lian_landau_2020}.

Despite efforts to bridge the gap between the two limits and provide a more general theory \cite{janecek_two-dimensional_2013,lian_open_2021,wang_narrow_2022,wang_revisiting_2023,kolar_hofstadter_2024}, a detailed description of the crossover is lacking and an explicit finite-field Wannier construction that reveals the flux dependence of the effective tight-binding parameters remains incomplete. This motivates us to study to what extent the conventional Hofstadter model is a reliable description of two-dimensional electrons at large magnetic flux, and in which limits it is recovered. Band topology provides a natural bridge between the two descriptions because Chern numbers remain invariant under continuous changes of representation as long as the relevant gaps remain open \cite{thouless_quantized_1982,macdonald_landau-level_1983}.

\begin{figure*}[ht!]
    \includegraphics[width=\textwidth]{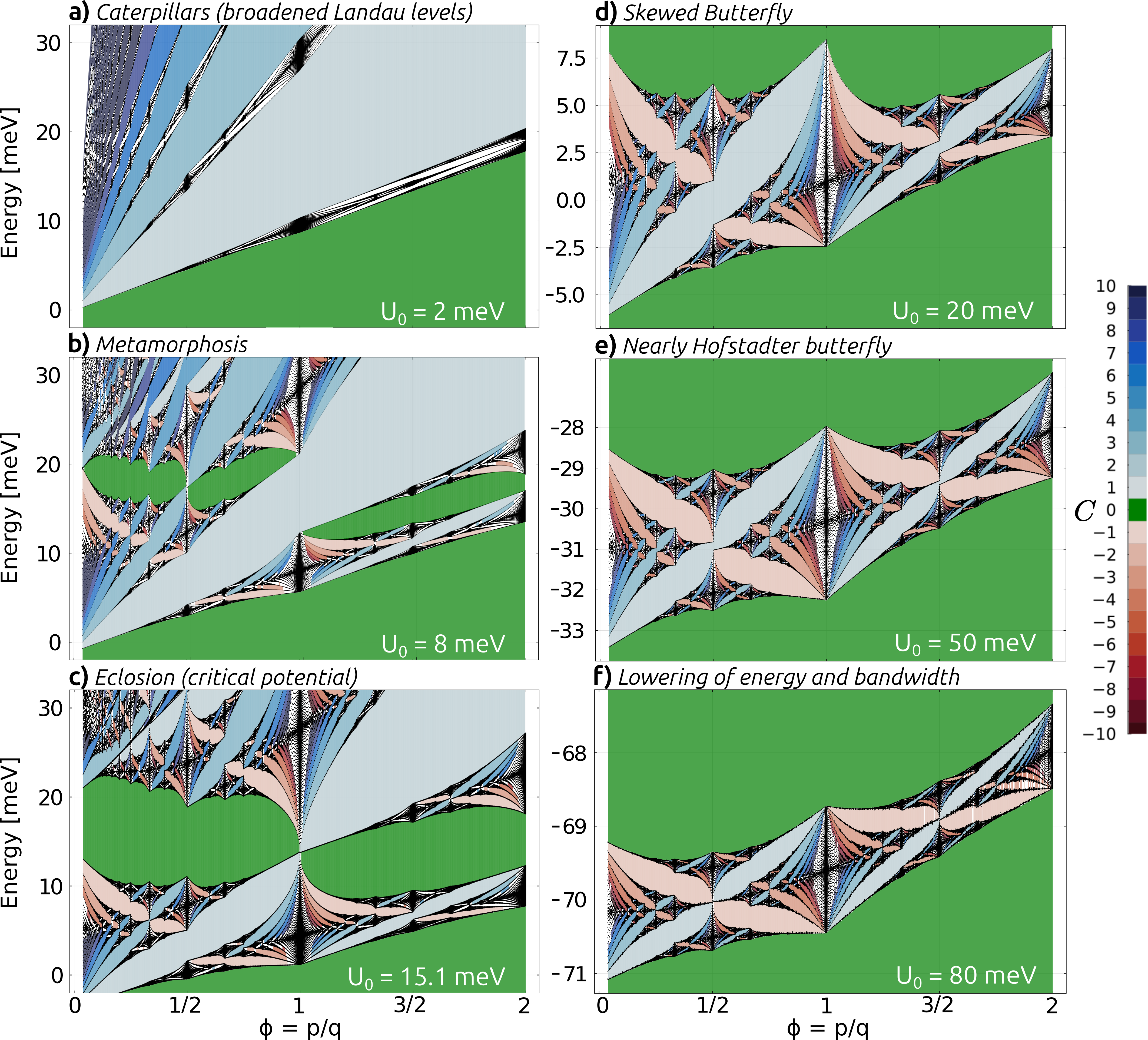}
    \caption{\label{fig:spectra} Low-energy spectra for increasing periodic-potential strengths $U_0=2$, $8$, $15.1$, $20$, $50$, and $80\,\mathrm{meV}$, in reading order. Black dots denote the physical states, and the color of each gap indicates the cumulative Chern number of the bands below it. Panels \textbf{(a)}--\textbf{(c)} show the gradual broadening and hybridization of the Landau levels into a lowest Hofstadter-like band separated from the rest of the spectrum. Panel \textbf{(c)} is at the critical potential strength, above which a topologically trivial gap remains open at all fluxes. Panels \textbf{(d)}--\textbf{(f)} focus on this isolated lowest band.}
\end{figure*}

Based on a numerical exact diagonalization, we uncover the precise evolution from Landau levels to butterflies, as presented in our main Fig.~\ref{fig:spectra}. For a weak potential, Landau levels are broadened to form `caterpillars' that host inverse butterfly spectra. A series of topological transitions at fluxes $\phi = 1/q$ isolate a trivial low-energy band, culminating in an eclosion where the butterfly emerges as a distinct topologically trivial band for sufficiently strong potential. That band remains different from the Hofstadter picture: both the amplitude of the hopping parameters, and their complex phases are different, leading to a skewed butterfly with a different Chern number sequence than in an extended Hofstadter model. These results reflect the richness of the problem where the magnetic field and the periodic potential are both strong.

The remainder of this paper is structured as follows. We study noninteracting spinless electrons in a square cosine potential as a minimal setting in which to address this problem. We diagonalize the continuum Hamiltonian in a magnetic-translation-adapted Landau-level basis, as described in Sec.~\ref{sec:hamil}. In Sec.~\ref{sec:snt}, we show how Landau "caterpillars" gradually evolve into Hofstadter butterflies within an isolated, topologically trivial lowest band, and we develop an analytical description of the resulting topological structure and principal transitions. In Sec.~\ref{sec:tb}, we construct a finite-field Wannier basis, extract the magnetic-field-dependent tight-binding parameters, and compare the resulting extended tight-binding model with the exact continuum spectrum. In Sec.~\ref{sec:diagonalt}, we assess the validity of the Hofstadter description and present evidence that the conventional straight-line Peierls contribution to the diagonal hopping need not be dominant at larger fluxes.

\section{\label{sec:hamil}A magnetic field and a periodic potential}

To calculate the electronic spectrum in the presence of a magnetic field and a periodic potential, we consider an infinite system of free, noninteracting, spinless electrons in two dimensions. The electrons are subject to a perpendicular magnetic field $B$ and a periodic potential $U(x,y)$. The well-known single-particle Hamiltonian is
\begin{align}\label{eq:hamiltonian_continuum}
    H = \frac{(\bm{p}-e\bm{A})^2}{2m_e} + U(x,y)
\end{align}
where $e$ and $m_e$ are the electron charge and mass, respectively. The momentum operator is $\bm{p}=-i\hbar\bm{\nabla}$, $\bm{A}$ is the vector potential, and $\hbar$ is the reduced Planck constant.

To set the stage, and to provide clarity, we will now recap the main ingredients required to solve the above Hamiltonian. The goal is to find, for any potential and magnetic field strength the full spectrum of electronic eigenstates.

\begin{figure*}
    \includegraphics{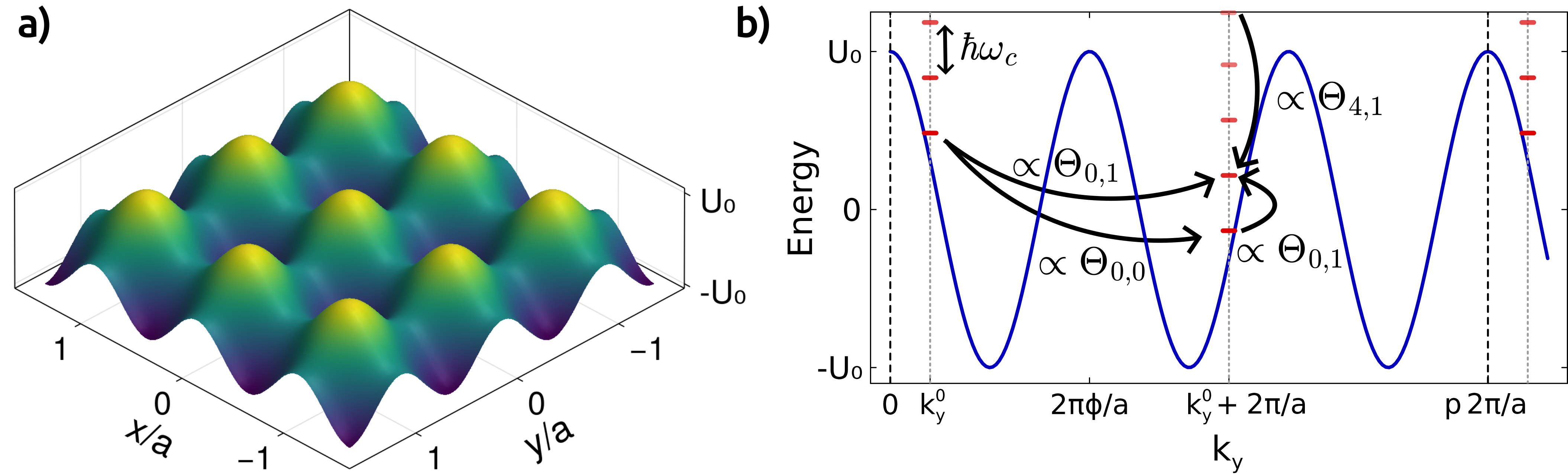}
    \caption{\label{fig:lattice} \textbf{(a)} Periodic potential $U(x,y)$. \textbf{(b)} Effective one-dimensional hopping chain in momentum $k_y$ (or position $x$) for $k_y^0=0$, obtained from the Hamiltonian in Eq.~\eqref{eq:hamiltonian_1d}. The blue curve is the on-site potential induced by $U(x,y)$, and the red lines are the available energy states.}
\end{figure*}

\subsection{Recap: Landau level basis}
In the absence of a periodic potential, the above Hamiltonian of Eq.~\eqref{eq:hamiltonian_continuum} leads to the formation of Landau levels \cite{macdonald_introduction_1994}. In this work, we choose to work in the Landau gauge
\begin{align}
    \bm{A}(\bm{r})=\bm{A}(x,y) =[0,Bx,0],
\end{align}
for which $\bm{\nabla} \times {\bm A} = B \hat{z}$. The eigenstates of the Hamiltonian form the so-called Landau-level basis
\begin{align}\label{eq:landau_basis}
\ket{LL_{n,k_y}(x,y)} &\propto H_n(\xi_x)\,e^{\nicefrac{-\xi_x^2}{2}}\,e^{ik_y y}\,,\nonumber\\
\xi_x &= \tfrac{x}{l_B}-k_yl_B,
\end{align}
where $l_B=\sqrt{\hbar/(eB)}$ is the magnetic length and $H_n(x)=(-1)^n e^{x^2}\frac{\partial^n}{\partial x^n}e^{-x^2}$ are the Hermite polynomials. 

The quantum numbers are the Landau-level index $n\in\mathbb{N}_0$ and the free-particle momentum $k_y\in\mathbb{R}$. The momentum $k_y$ fixes the guiding center of the state at $\langle x \rangle=k_y l_B^2$. The Landau-level wave functions in Eq.~\eqref{eq:landau_basis} are orthonormal, $\langle LL_{n',k_y'}|LL_{n,k_y}\rangle=\delta_{nn'}\delta_{k_yk_y'}$, and, in the absence of a periodic potential, satisfy
\begin{align}
    \frac{(\bm{p}-e\bm{A})^2}{2m_e} \,\ket{LL_{n,k_y}} = \varepsilon_n\,\ket{LL_{n,k_y}}
\end{align}
where the Landau level energy is given by
\begin{align}
    \varepsilon_n=\hbar\omega_c\left(n+\tfrac{1}{2}\right)
    \label{Eq:LandauLevelEnergy}
\end{align}
and the cyclotron frequency is defined as
\begin{align}
    \omega_c=\frac{eB}{m_e}.
    \label{Eq:CyclotronFrequency}
\end{align}

\subsection{Recap: Adding the periodic potential}
We now introduce an isotropic periodic potential $U(x,y)$. Any two-dimensional lattice is defined by a set of translation operators $T_{x}^a$ and $T_{y}^a$. In the absence of a magnetic field, these translation operators commute with each other and with the Hamiltonian. For simplicity, we consider a square lattice with lattice constant $a$, so that $T_x^a U(x,y) = U(x+a,y) = U(x,y)$ and $T_y^a U(x,y) \equiv U(x,y+a) = U(x,y)$.

In the presence of a magnetic field, these ordinary translation operators no longer commute with the Hamiltonian. Instead, there exist modified \textit{magnetic translation operators} $\Tilde{T}$, which do commute with the Hamiltonian and can therefore correspond to relevant quantum numbers. To construct them, we need to counteract the gauge dependence of the Hamiltonian, such that the magnetic translation operator in $x$ satisfies
\begin{align}
    \Tilde{T}_{x}^a = e^{i\tfrac{e}{\hbar}\chi_x^a(\bm{r})}\,T_{x}^a\,,\quad \bm{A}(\bm{r}+\hat{x}a)=\bm{A}(\bm{r}) + \bm{\nabla}\chi_x^a(\bm{r})\,
\end{align}
and similarly for $\Tilde{T}_y$. Interestingly, the two magnetic translations, in general, do not commute,
\begin{align}
    \Tilde{T}_{x}^a\,\Tilde{T}_{y}^a = e^{-i2\pi\phi} \Tilde{T}_{y}^a\,\Tilde{T}_{x}^a
\end{align}
where 
\begin{align}
    \phi = \frac{Ba^2}{\Phi_0}
    =\frac{a^2}{2\pi l_B^2}
    \label{Eq:FluxQuanta}
\end{align}
is the number of magnetic flux quanta $\Phi_0=h/e$ per unit cell. This quantity $\phi$ plays a fundamental role in the study of the electronic spectrum. For rational fluxes
\begin{align}
    \phi=p/q,
    \label{Eq:RationalFlux}
\end{align}
we can enlarge the lattice unit cell to a \textit{magnetic unit cell} containing $q$ lattice unit cells. Translations by the magnetic-unit-cell vectors then commute with each other and with the Hamiltonian. In the Landau gauge, we can choose $\chi_y^a=0$ and $\chi_x^a=Bay$, so a magnetic translation along $x$ acquires the phase $2\pi\phi y/a$. This results in a magnetic unit cell, which extends to $q$ lattice unit cells in a row in the $x$ direction.

Note that this means that at a rational flux, each Hamiltonian eigenstate must be simultaneously an eigenstates of the operators $T^a_y$ and $T^{qa}_x$. We will get back to this point later when discussing the quantum numbers of the eigenstates.

If $\phi$ is irrational, no finite enlargement of the lattice unit cell can make the magnetic translations commute, and hence there is no finite magnetic unit cell or ordinary magnetic Bloch-band decomposition. The irrational case is therefore commonly approached through rational approximations $p_n/q_n\to \phi$ whose magnetic unit cells grow without bound. Irrational fluxes can therefore be viewed as providing a continuous spectral extension between nearby rational-flux calculations, although individual magnetic subbands cannot generally be continued one to one because their count changes with $q_n$. In the Hofstadter model, for example, the spectrum is bounded by a Cantor set \cite{rauh_bloch_1974,hofstadter_energy_1976}.

Although the infinite system contains infinitely many states, the number of states per Landau level and per unit cell is well-defined and equals the number of magnetic flux quanta per unit cell, $\phi$.

\subsection{Numerical basis}

In the remainder of this paper we restrict ourselves to rational fluxes following Eq.~\eqref{Eq:RationalFlux}. In this case, provided $q$ is not too large, we can numerically diagonalize the full Hamiltonian of Eq.~\eqref{eq:hamiltonian_continuum}. Practically, we construct the Hamiltonian matrix elements
\begin{eqnarray}
    \label{eq:hamil1}
    \Braket{LL_{n',k_y'}|H|LL_{n,k_y}}
\end{eqnarray}
explicitly. We consider the simple two-dimensional cosine periodic potential
\begin{eqnarray}
    U(x,y) = U_0\left(
    \cos(2\pi \tfrac{x}{a})  + \cos(2\pi \tfrac{y}{a}) \right)\,
\end{eqnarray}
shown in Fig.~\ref{fig:lattice}(a). The precise expressions for the matrix elements are given in Appendix~\ref{ap:matrix_el}. 

The periodic potential in general mixes states with different Landau level index $n$. However, the $x$-dependent term does not mix states with different $k_y$, just as in the potential-free Hamiltonian. By contrast, the $y$-dependent term mixes states whose momenta $k_y$ differ by $2\pi/a$. We therefore decompose the momentum as
\begin{align}
    k_y=k_y^0+\frac{2\pi}{a}M,
    \label{Eq:kyDecomposition}
\end{align}
where $k_y^0\in[0,2\pi/a)$ and $M\in\mathbb{Z}$. The problem can then be interpreted as a family of one-dimensional hopping chains in momentum space, one for each $k_y^0$, as shown in Fig.~\ref{fig:lattice}(b). In this analogy, the $x$-dependent term of $U$ produces a spatially varying on-site potential with period $2\pi \phi / a$, while the hopping step is $\Delta k_y = 2\pi/a$. 

After $p$ number of hops in the same direction of $k_y$, equivalent to $q$ periods of the $x$-dependent term potential, the chain reaches a site with the same on-site potential as the starting site. This defines a flux-dependent unit cell of length $p\, \tfrac{2\pi}{a}=q\, \frac{2\pi}{a} \phi$ in the one-dimensional chain. We therefore decompose the $k_y$-momentum further,
\begin{eqnarray}
    k_y = k_y^0+K_y\,\frac{2\pi}{a}p + m\,\frac{2\pi}{a}
\end{eqnarray}
where $K_y\in\mathbb{Z}$ and $m\in\{0,\ldots,p-1\}$. Because the Hamiltonian is periodic in $K_y$, we can introduce $Y\in[0,2\pi)$, the Fourier conjugate of $K_y$, which further block-diagonalizes the Hamiltonian. 

The quantities $Y/(qa)$ and $k_y^0$ are good momentum quantum numbers of the magnetic translation group \cite{zak_magnetic_1964}. Indeed, using the convention $(T_\mu^d\psi)(\bm r)=\psi(\bm r-d\hat{\bm e}_\mu)$, the states in Eq.~\eqref{eq:psi_basis} satisfy
\begin{align}
    &\tilde{T}_x^{qa}\ket{\psi_{n,m,X,Y}}
        =e^{-iY}\ket{\psi_{n,m,X,Y}},\\
    &\tilde{T}_y^{a}\ket{\psi_{n,m,X,Y}}
        =e^{-ik_y^0a}\ket{\psi_{n,m,X,Y}}
         =e^{-iX/q}\ket{\psi_{n,m,X,Y}}. \nonumber
\end{align}
For the remainder of the calculation, it is convenient to use the dimensionless parameters $Y$ and $X=k_y^0qa\in[0,2\pi q)$. At the same time, these parameters determine the guiding-center due to position-momentum coupling in the Landau levels, as will be discussed later.

We have thus constructed our numerical basis, described by the creation operator $\hat{\psi}^\dagger_{n,m,X,Y}$ that creates a superposition of Landau-level states when acting on the vacuum
\begin{eqnarray}\label{eq:psi_basis}
    \hat{\psi}^\dagger_{n,m,X,Y}\ket{0}=\sum_{K_y=-\infty}^\infty e^{-iK_yY}\ket{LL_{n,k_y(X,K_y,m)}}
\end{eqnarray}
and the conjugate annihilation operator $\hat{\psi}_{n,m,X,Y}$. To normalize $\ket{\psi_{nmXY}}$ in practical calculations, we replace the infinite sum over $K_y$ by $\frac{1}{\sqrt{2K_y^{\max}+1}}\sum_{K_y=-K_y^{\max}}^{K_y^{\max}}$ and take the limit $K_y^{\max}\to\infty$ at the end. Because $K_y$ labels magnetic unit cells along $x$, small values of $K_y^{\max}$ are sufficient when calculating properties within a single magnetic unit cell.

\subsection{Hamiltonian}

The electronic eigenstates carry well-defined quantum numbers $X,Y$. This numerical basis allows us to form, for each $X,Y$, a Hamiltonian matrix with Landau level index $n \in \{ 0, 1, \ldots \}$ and the orbital number $m \in \{ 0, \ldots, p-1 \}$. This remaining matrix is solved numerically. 
Starting from Eq.~\eqref{eq:hamil1}, we can now rewrite the full Hamiltonian in terms of the basis of Eq.~\eqref{eq:psi_basis} as
\begin{widetext}
\begin{eqnarray}\label{eq:hamiltonian_1d}
    \hat{H}(\phi)=\int_0^{2\pi q}dX \int_0^{2\pi}\frac{dY}{2\pi} && \sum_{n,n'=0}^\infty \sum_{m=0}^{p-1}\bigg\{ \left(\varepsilon_n \,\delta_{n,n'} +  U_0\,\Theta_{n,n'}(\phi)\,\cos\left[\tfrac{X}{p}+m\tfrac{2\pi}{\phi}+\tfrac{\pi}{2}|n'-n|\right]\right)\; \hat{\psi}^\dagger_{n,m}\hat{\psi}_{n',m} \nonumber \\
    && + \frac{U_0}{2}\Theta_{n,n'}\,(-1)^{\tfrac{n-n'+|n'-n|}{2}}\left(e^{i\frac{Y}{p}}\;\hat{\psi}^\dagger_{n,m}\hat{\psi}_{n',m-1} + (-1)^{n'-n}\;e^{-i\frac{Y}{p}}\;\hat{\psi}^\dagger_{n,m}\hat{\psi}_{n',m+1} \right)\bigg\}\,.
\end{eqnarray}
\end{widetext}
Here and below, the $X$ and $Y$ indices of $\psi^\dagger$ are implicit. The symmetric form factors $\Theta_{n,n'}(\phi)=\Theta_{n',n}(\phi)$ involve generalized Laguerre polynomials and determine the flux-dependent Landau-level broadening; they are derived in Appendix~\ref{ap:matrix_el} and are given by:
\begin{align}
\Theta_{n,n'}(\phi)
={}& e^{-\frac{\pi}{2\phi}}
\sqrt{\tfrac{\min(n,n')!}{\max(n,n')!}}
\left(\frac{\pi}{\phi}\right)^{\frac{|n'-n|}{2}}
\nonumber\\
&\times L^{|n'-n|}_{\min(n,n')}
\left(\frac{\pi}{\phi}\right)
\label{eq:theta_matrix_elements}
\end{align}
where $L^{(\alpha)}_n(x)=\frac{x^{-\alpha}e^x}{n!}\frac{\partial^n}{\partial x^n}\left(e^{-x}x^{n+\alpha}\right)=\sum_{k=0}^n\tbinom{n+\alpha}{\alpha+k}\tfrac{(-x)^k}{k!}$ are the generalized Laguerre polynomials. The resulting Hamiltonian matrix has elements $\mathcal{H}_{nn'}^m(\phi,X,Y)$, defined through $H(\phi)=\int_{X,Y}\sum_{nn'm}\mathcal{H}_{nn'}^m(\phi,X,Y)$. Throughout the rest of this work, rows and columns follow the basis ordering $\{\hat{\psi}_{0,0},\hat{\psi}_{0,1},\dots,\hat{\psi}_{0,p-1},\hat{\psi}_{1,0},\dots\}$.

\subsection{Visualizing the quantum numbers $X$, $Y$}

\begin{figure}[h]
\includegraphics[width=\linewidth]{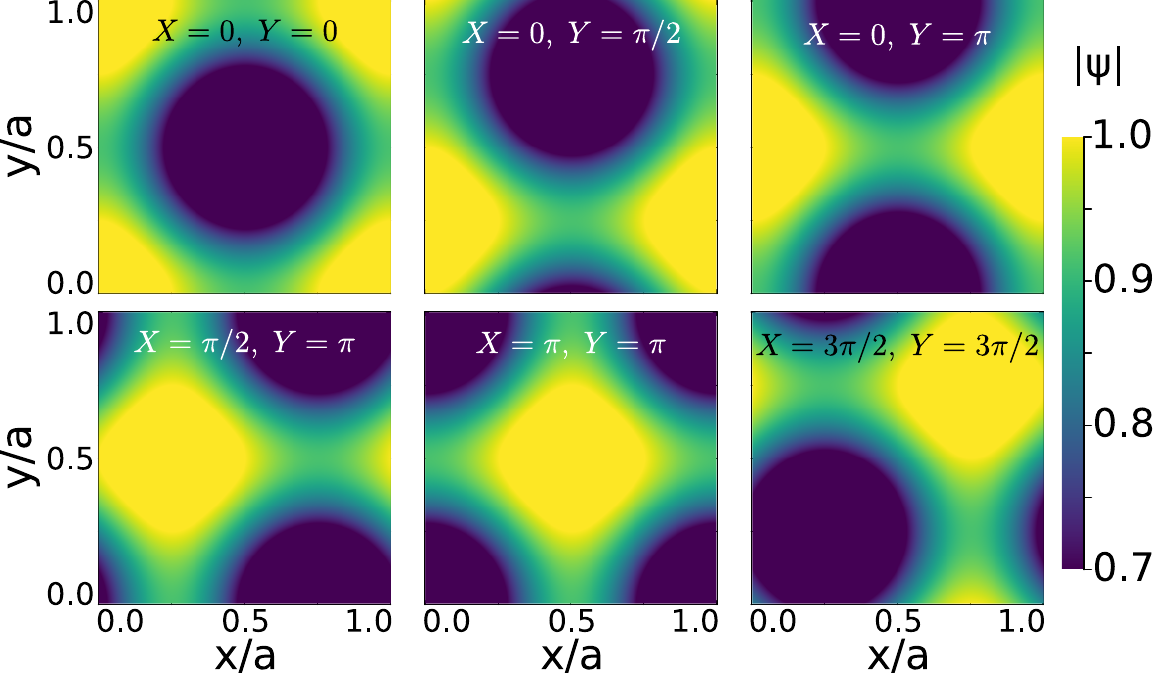}
\caption{\label{fig:jacobi_theta} Density of an unmixed lowest-Landau-level exact analytical solution at $\phi=1$ for a few selected eigenstates with different momentum quantum numbers $X$ and $Y$, demonstrating that the these quantum numbers also relate to a real space guiding center.}
\end{figure}

To clarify the role of the quantum numbers $X$ and $Y$, we focus on $p=1$ and consider the lowest Landau level ($n=0$) in isolation. The Hamiltonian then takes the diagonal form
\begin{eqnarray}
    \hat{H}(1/q)=&&\int_{0}^{2\pi q} dX\int_0^{2\pi}\frac{dY}{2\pi}\\
    &&\qquad\psi^\dagger\psi \left( \varepsilon_0 + U_0 e^{-q\frac{\pi}{2}} \left(\cos X+\cos Y\right) \right).\nonumber
\end{eqnarray}
Starting from Eq.~\eqref{eq:psi_basis}, we can calculate the wave function associated with the creation operator $\psi^\dagger$ analytically:
\begin{eqnarray}
    \ket{\psi_{XY}}\propto e^{-\frac{\pi}{q}\frac{(y-\tfrac{a}{2\pi}Y)^2}{a^2}}\;\vartheta\left(-\frac{(z-\tfrac{a}{2\pi}Z)}{aq};\,\frac{i}{q}\right)
\end{eqnarray}
where $\vartheta(z;\tau)=\sum_{n=-\infty}^\infty \exp(i\pi \tau n^2+i2\pi z n)$ is the doubly periodic Jacobi theta function, $z=x+iy$, and $Z=X+iY$. The momentum-related quantum numbers thus specify the guiding-center position $(x_c,y_c)=\tfrac{a}{2\pi}(X,Y)$ within the magnetic unit cell. This also explains the $X$ and $Y$ dependence of the energy: it depends on where the wave-function center lies relative to the periodic potential. At lower flux (larger $q$), the wave function is more extended and its bandwidth is smaller. The quantity $1/q$ is both the ratio of the Jacobi-theta-function periods and the aspect ratio of the magnetic unit cell. Representative densities are shown in Fig.~\ref{fig:jacobi_theta}, which visualizes how indeed $(X,Y)$ act as guiding centers for the Bloch functions within the unit cell.

For $q>1$, the same construction remains exact within the isolated lowest Landau level -- the magnetic unit cell extends over $q$ ordinary unit cells, $X$ spans this enlarged cell, and the guiding-center interpretation continues to hold. Increasing $q$ increases the magnetic length and suppresses the bandwidth exponentially through the factor $e^{-\pi q/2}$. This simple picture applies to any integer $q$ when $p=1$. For $p>1$, however, there are $p$ coupled states at each magnetic Bloch momentum, so $X$ and $Y$ can no longer be identified with the center of a single eigenstate. Landau-level mixing likewise modifies the analytic wave function, although the guiding-center picture remains useful qualitatively in the weak-mixing regime.

\subsection{Energy and length scales}

The problem is controlled by three energy scales: the periodic-potential amplitude $U_0$, the cyclotron energy $\hbar\omega_c$, and the kinetic Umklapp scale at the edge of the zero-field Brillouin zone,
\begin{align}
    E_U=\frac{\hbar^2}{2m_e}\left(\frac{\pi}{a}\right)^2.
\end{align}
At fixed flux $\phi$, the cyclotron energy can be written as
\begin{align}
    \hbar\omega_c=\frac{2\pi\hbar^2}{m_ea^2}\phi=\frac{4\phi}{\pi}E_U.
\end{align}
Thus, both intrinsic kinetic scales vary as $a^{-2}$, and the spectra depend on $a$ only through the overall energy scale and the dimensionless ratio $U_0/E_U$. Throughout this work, we use $a=5\,\mathrm{nm}$, representative of a short moir\'e period, for which $E_U\simeq15.1\,\mathrm{meV}$. Although our square potential is not intended to model the geometry of a specific moir\'e material, this choice places the transitions in a realistic moir\'e energy range. Results for another lattice constant $a'$ follow by rescaling
\begin{align}
    U_0(a')=U_0(5\,\mathrm{nm})\left(\frac{5\,\mathrm{nm}}{a'}\right)^2
\end{align}
at fixed $U_0/E_U$ and $\phi$. The corresponding magnetic field scales by the same factor.

The relevant length scales are $a$ and the magnetic length $l_B$. Their ratio is fixed entirely by the flux,
\begin{align}
    \frac{l_B}{a}=\frac{1}{\sqrt{2\pi\phi}},
\end{align}
while the displacement between neighboring sites of the effective one-dimensional chain is $(2\pi/a)l_B^2=a/\phi$. Increasing $\phi$ therefore reduces the size of a Landau orbital relative to the unit cell and makes the electron more sensitive to the local potential. This is the same crossover that was seen for the energy by the linear increase of $\hbar \omega_c/E_U=4\phi/\pi$.

Therefore, all physical results are dependent on only two relevant parameters: the dimensionless flux per unit cell $\phi$, and the ratio of the periodic potential strength and the Umklapp energy $U_0/E_U$.

\section{\label{sec:snt}Results: Spectra and topology}

Based on the numerical basis presented in previous section, we developed a code \cite{github_landau_butterflies} that allows for an efficient calculation of the full spectrum by exactly diagonalizing the Hamiltonian of Eq.~\eqref{eq:hamiltonian_1d}. 
Practically, this requires a cut-off in the number of Landau levels included. The accuracy of the results is therefore controlled primarily by the number $N_{\text{LL}}$ of Landau levels retained. Convergence of the low-energy spectrum requires a larger $N_{\text{LL}}$ at smaller flux $\phi$ and larger $U_0$, because the potential-induced mixing becomes stronger relative to the Landau-level spacing. We use the estimate
\begin{align}
N_{\text{LL}}
= \frac{600}{\phi}\frac{U_0}{1\,\text{eV}}
\end{align}
and round it up to the nearest integer.

To characterize the topology of the resulting spectrum, we track the cumulative Chern number $C$, defined as the sum of the Chern numbers of all bands below a given spectral gap. We calculate $C$ numerically using the St\v{r}eda formula \cite{streda_theory_1982},
\begin{align}
C = \left.\frac{\partial \nu}{\partial \phi}\right|_{E_F},
\end{align}
where $\nu$ is the integrated number of occupied states per ordinary unit cell and the Fermi energy $E_F$ is held inside the gap. Practically, this is done by identifying straight lines on a Wannier plot \cite{wannier_result_1978}.

\subsection{Spectrum of states}

In Fig.~\ref{fig:spectra} we show how the full spectrum continuously evolves from broadened Landau levels to a Hofstadter butterfly-like shape, by increasing the potential strength $U_0$. 

For weak $U_0$, the periodic potential broadens the Landau levels into magnetic subbands. When Landau-level mixing is neglected, each isolated Landau level develops an internal structure of an `inverse butterfly'. In contrast to the conventional tight-binding butterfly, whose pattern is periodic in $\phi$, the inverse-butterfly structure is periodic in $1/\phi$, apart from its flux-dependent overall energy shift and bandwidth \cite{claro_magnetic_1979,thouless_quantized_1982}. The coupling between Landau levels further open gaps at spectral crossings, including the Dirac points that remain gapless when each Landau level is treated in isolation. This weak-potential regime is shown in Fig.~\ref{fig:spectra}(a). Inspired by the `butterfly' analogy, the spectrum now resembles long caterpillars waiting to grow further.

And as $U_0$ is increased, the gaps within caterpillars grow. While the gaps \textit{between} Landau levels carry consecutive Chern numbers $C = 1, 2, 3, \ldots$, the gaps \textit{within} the Landau levels have a richer structure. In particular, a cumulative trivial gap with $C=0$ seems to grow the most, as can be seen in Fig.~\ref{fig:spectra}(b), where gaps are colored green. 
At fluxes $\phi=1/q$, the relevant gaps close and reopen as $U_0$ increases past certain values, transferring a Chern number between the adjacent bands such that the reopened gap has $C=0$. We analyze this mechanism in detail in Sec.~\ref{Sec:TopoTransitions}, and Fig.~\ref{fig:spectra}(b) shows such a topological gap closing at $\phi = 1/2$.

With further increasing $U_0$, the trivial gaps associated with different Landau levels connect across the varying fluxes and form a single expanding gap that separates the lowest band from the rest of the spectrum.
At the critical potential strength $U^c$, shown in Fig.~\ref{fig:spectra}(c), this lowest $C=0$ gap opens throughout all fluxes. 
From now on, a trivial band is separated from the rest of the spectrum: this is the caterpillar's eclosion, its emergence as a butterfly. The trivial band hosts exactly one electronic state per $(X,Y)$ quantum numbers per lattice unit cell.

This is one of our key results: in order to recover the Hofstadter butterfly, the first step is to create this trivial band disconnected from the rest of the spectrum. 
At still larger $U_0$, additional trivial gaps appear at higher energies, producing further isolated bands with zero net Chern number. 

Now that the eclosion has occurred and an isolated butterfly has emerged, we note that the resulting butterfly is still different from Hofstadter's butterfly.
The spectrum of the lowest isolated band displays a fractal subband structure and a sequence of Chern-labeled gaps reminiscent of the Hofstadter butterfly. It is not, however, a rigid copy of the conventional butterfly -- both its central energy and its total bandwidth depend on $\phi$, and it contains additional gaps that are absent from the nearest-neighbor Hofstadter model. As $U_0$ increases further, the band shifts to lower energy and becomes narrower, while these additional gaps tend to close. In Fig.~\eqref{fig:spectra}(d)-(f) we show how this butterfly evolves. We quantify this evolution, and the differences with Hofstadter physics, in Sec.~\ref{sec:tb} by constructing an effective tight-binding model with flux-dependent parameters.

Given that we started with Landau levels, it is interesting to notice that the final butterfly has branches reminiscent of Landau levels, something which was already noticed by Hofstadter \cite{hofstadter_energy_1976}. However, the Landau levels in the bottom left of Fig.~\eqref{fig:spectra}(f) are not smoothly connected to the original Landau levels, and are, in fact, complicated superpositions of the original ones.

The overall spectral evolution is qualitatively consistent with the results of Janecek \textit{et al.}, who studied a square array of localized potential wells and similarly followed the continuum spectrum departing from the Landau-level regime \cite{janecek_two-dimensional_2013}. Their calculation solves the continuum Schr\"odinger equation in real space using magnetic Bloch boundary conditions and a diffusion-based numerical method. By contrast, our formulation uses an explicit Landau-level basis, making the composition and mixing of the Landau levels particularly transparent. The corresponding eigenvectors are used below to analyze the topological transitions, construct Wannier states, and extract the flux-dependent hopping parameters of the effective tight-binding model.

\subsection{Topological phase transitions at $\phi=1/q$}
\label{Sec:TopoTransitions}

The complete separation of the trivial band requires a series of topological transitions whereby a nontrivial gap $(C>0)$ closes and reappears as trivial $(C=0)$. These transitions occur at rational fluxes $\phi = 1/q$, and we can make an analytical approximation to describe how this transition works. 

We start at flux $\phi=1$, arguably the simplest case. Let us focus on the two lowest Landau levels and construct the $2\times 2$ Hamiltonian matrix
\begin{eqnarray}
    \mathcal{H} &=& \mathbbm{1}\,\left( \tfrac{\varepsilon_0+\varepsilon_1}{2} + U_0\,e^{-\tfrac{\pi}{2}}\,(1-\tfrac{\pi}{2})\,(\cos X + \cos Y)\right)\nonumber\\
    &-& \sigma^x\, U_0\,e^{-\tfrac{\pi}{2}}\sqrt{\pi}\sin X + \sigma^y\, U_0\,e^{-\tfrac{\pi}{2}}\sqrt{\pi}\sin Y\nonumber\\
    &+&\sigma^z\,\left( \tfrac{\varepsilon_0-\varepsilon_1}{2} + U_0\,e^{-\tfrac{\pi}{2}}\tfrac{\pi}{2}\,(\cos X + \cos Y)\right).
    \label{Eq:HamiltonianFlux1}
\end{eqnarray}
At the critical potential 
\begin{align}
    U_0^c=\frac{\varepsilon_1-\varepsilon_0}{2\pi}e^{\tfrac{\pi}{2}}
    = \frac{\hbar\omega_c}{2\pi}e^{\tfrac{\pi}{2}},
\end{align} 
the spectrum is gapless and has a Dirac cone at $X=Y=0$. Taken at plain value, we can calculate the Berry curvature of the eigenstates of Eq.~\eqref{Eq:HamiltonianFlux1} and integrate it numerically over the parameter space $(X,Y)$. Below the critical potential, the lower and upper bands of the two-band Hamiltonian have Chern numbers $(0,0)$, respectively. Above it, they have Chern numbers $(-1,1)$. This shows that the critical potential represents a topological transition.

This two-level Hamiltonian, however, is expressed in the topologically nontrivial basis defined in Eq.~\eqref{eq:psi_basis}. The \textit{actual} topology of the electronic eigenstates should take into account the topology of the Landau level basis. We found that the numerical basis we use is especially useful to calculate the Chern number of Landau levels. In fact, the Chern number of each basis state is
\begin{eqnarray}
    C_{n,m}(\phi)=i\oint \frac{d\bm{R}}{2\pi}\cdot \braket{\psi_{nmXY}|\partial_{\bm{R}}\psi_{nmXY}} = \frac{1}{p}
    \label{Eq:ChernNumberBasis}
\end{eqnarray}
so that summing over the $p$ bands within a Landau level gives $C=1$. Here, $\bm{R}=(X,Y)$, and the integral follows the boundary of the parameter space. Details are given in Appendix~\ref{ap:chern_landau}. This result allows us to explicitly calculate the Chern numbers based on the eigenstates; and not just using the aforementioned procedure based on Středa's formula. The explicit derivation of Eq.~\eqref{Eq:ChernNumberBasis} has not been reported in the literature so far, though a similar construction has been \cite{usov1988theory}.
At $\phi=1$, where $p=1$, adding the basis-state contribution $(1,1)$ gives total Chern numbers $(1,1)$ below and $(0,2)$ above the critical potential, in agreement with our numerical result obtained from the St\v{r}eda formula. Note that in Fig.~\ref{fig:spectra}(c), this topological transition point is visualized.

The description of the topological transition at $\phi = 1$ can be extended to fluxes $\phi=1/q$. The two-band Hamiltonian formed by the $q$th and $(q-1)$th Landau levels is
\begin{eqnarray}
    \mathcal{H} &&= \mathbbm{1}\,\left( \tfrac{\varepsilon_{q-1}+\varepsilon_q}{2} + U_0\,\tfrac{\Theta_{q-1,q-1}+\Theta_{q,q}}{2}\,(\cos X + \cos Y)\right)\nonumber\\
    &&- \sigma^x\, U_0\Theta_{q-1,q}\sin X + \sigma^y\, U_0\Theta_{q-1,q}\sin Y \\
    +&&\sigma^z\,\left( \tfrac{\varepsilon_{q-1}-\varepsilon_q}{2} + U_0\,\tfrac{\Theta_{q-1,q-1}-\Theta_{q,q}}{2}\,(\cos X + \cos Y)\right). \nonumber
\end{eqnarray}
This Hamiltonian is a variation of the Qi-Wu-Zhang model \cite{qi_topological_2006,qi_topological_2008}, and it produces a lower-band Chern number
\begin{eqnarray}
    C=\Bigg\{\begin{array}{ll}
        -1 & \quad \text{if}\quad \bigg|\frac{\varepsilon_{q}-\varepsilon_{q-1}}{U_0(\Theta_{q,q}-\Theta_{q-1,q-1})}\bigg|<2 \\
        0 & \quad \text{otherwise.}
    \end{array}
\end{eqnarray}
This result agrees with the $\phi=1$ case above. The Dirac cones occur at $X=Y=0$ for odd $q$ and at $X=Y=\pi$ for even $q$ (modulo $2\pi$). They are equivalent in each $2\pi$ period of $X$. Therefore, for $q>1$, the enlarged magnetic unit cell produces a larger parameter space and effectively multiplies the Chern number by $q$ -- the number of Dirac cones in the Brillouin zone. Because all Dirac points close and reopen the gap simultaneously, 
each $q$ has a unique critical potential
\begin{eqnarray}\label{eq:ucrit_2b}
    U_0^c&&=\bigg|\frac{\hbar\omega_c}{2(\Theta_{q,q}-\Theta_{q-1,q-1})}\bigg|\,,
\end{eqnarray}
where a topological transition occurs from a $C=q$ gap to a topologically trivial $C=0$ gap. There are no gap closings at other fluxes during the separation of the lowest trivial band from the rest of the spectrum. This can be thought of as a consequence of the spectrum of the inverse butterfly in each Landau level being bounded by a Cantor set.

\begin{figure}
    \centering
    \includegraphics[width=\linewidth]{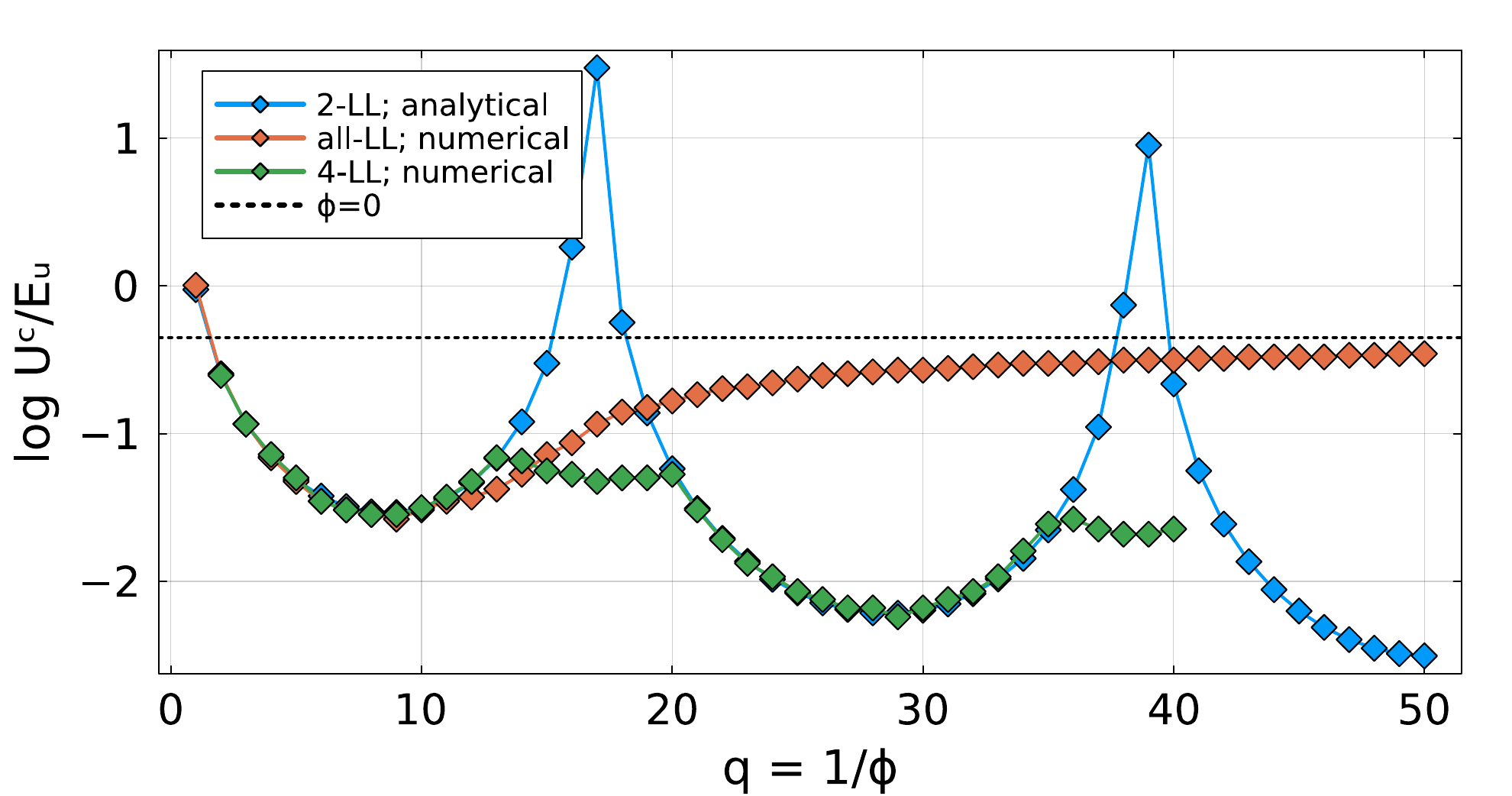}
    \caption{Critical potential strength $U_0^c$ on a logarithmic scale for a range of fluxes $\phi=1/q$. Above $U_0^c$, a topologically trivial gap separates the lowest Hofstadter-butterfly-like band from the rest of the spectrum.}
    \label{fig:Ucrit}
\end{figure}

The expression of Eq.~\eqref{eq:ucrit_2b} is an approximation, where we neglected the influence of other Landau levels. As a result, where Eq.~\eqref{eq:ucrit_2b} nearly diverges at certain rational values of the argument of the $\Theta_{n,n'}$ functions, the approximation is no longer reliable. Instead, when all relevant Landau levels are numerically included, the gaps at $\phi=1/q$ still close and reopen at a critical potential, but the divergent-like behavior disappears. Fig.~\ref{fig:Ucrit} compares the two results: notably, $\phi=1$ is the highest flux at which a transition is required to separate the lowest band. The two-Landau-level analytical result agrees quantitatively with the full numerical calculation only up to $q\sim10$. At lower flux, more than two Landau levels are required for an accurate representation.

Moreover, in Fig.~\ref{fig:Ucrit} we also compare the critical potential strengths $U^c$ with the potential needed to isolate a band in the absence of a magnetic field. Interestingly, the potential needed to separate the lowest energy trivial band is \textit{highest} at $\phi = 1$, then followed by $\phi = 0$, and only then other $\phi = 1/q$ with $q>1$. This means that once the lowest energy band is separated in the absence of a field, it will remain separated at any finite flux $\phi < 1$. However, since $U^c(\phi=1) > U^c(\phi=0)$, if the band is separated at zero flux does not guarantee that the trivial band remains separated at all fluxes.

\subsection{Gap opening at $\phi=2$}

\begin{figure}
\centering
    \includegraphics[width=\linewidth]{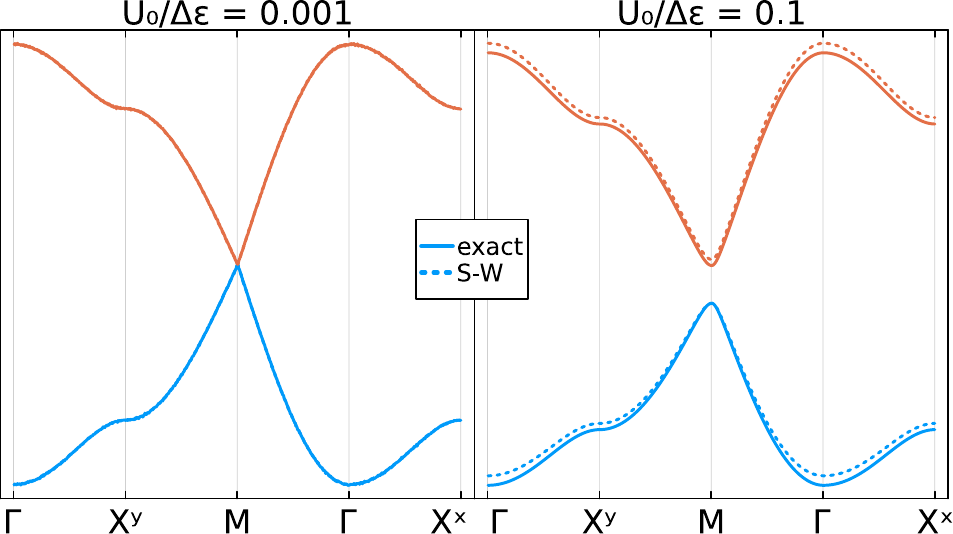}
    \caption{Energy dispersion of the lowest Landau level at $\phi=2$ along a selected high-symmetry path in the Brillouin zone. The solid line is the exact result obtained with ten Landau levels, and the dotted line is the two-Landau-level Schrieffer--Wolff approximation.}
    \label{fig:SW_gap}
\end{figure}

At fluxes above $\phi > 1$, the existence of a separate trivial band is no longer dependent on mixing between neighboring Landau levels. Instead, the resulting physics happens fully \textit{within} the lowest Landau level. 

In particular, let us look at flux $\phi=2$. In this case, the broadened spectrum of the isolated lowest Landau level (that is, without corrections induced by virtual excitations to higher Landau levels) has a Dirac cone at $X=Y=\pm\pi$, see Fig.~\ref{fig:SW_gap}, left. Mixing with higher Landau levels gaps out the Dirac cone. The gap is topologically trivial, which we confirmed numerically. 
To understand the gap opening analytically, we approximate the Hamiltonian by restricting it to the \textit{two} lowest Landau levels, that are weakly mixed. The Hamiltonian matrix is
\begin{equation}
    \mathcal{H}= \begin{bmatrix} \mathcal{H}_{n=0}^{\text{isol}} & \Delta^\dagger \\ \Delta & \mathcal{H}_{n=1}^{\text{isol}}\end{bmatrix}
\end{equation}
where
\begin{eqnarray}
    \mathcal{H}_{n}^{\text{isol}} = \mathbbm{1}\,\varepsilon_n &&+ \sigma^x \,U_0\Theta_{n,n}\cos\left(\tfrac{Y}{2}\right) \nonumber\\
    &&+ \sigma^z \, U_0\Theta_{n,n}\cos\left(\tfrac{X}{2}\right)
\end{eqnarray}
are the Hamiltonians of the isolated Landau levels. $\mathbbm{1}$ and $\sigma^j$ are the identity and Pauli matrices. The off-diagonal block
\begin{equation}
    \Delta = U_0\Theta_{1,0}\begin{bmatrix} -\sin\left(\tfrac{X}{2}\right) & i\sin\left(\tfrac{Y}{2}\right) \\ i\sin\left(\tfrac{Y}{2}\right) & \sin\left(\tfrac{X}{2}\right)  \end{bmatrix}
\end{equation}
mixes the two Landau levels. We perform a perturbative Schrieffer--Wolff transformation in $U_0/(\varepsilon_1-\varepsilon_0)$ and retain the leading correction. The resulting low-energy Hamiltonian for the lowest Landau level is
\begin{eqnarray}
    \mathcal{H}_{n=0}^*=  \mathcal{H}_{n=0}^{\text{isol}} -&& \mathbbm{1}\,\left(\Theta_{1,0}^2\tfrac{U_0^2}{\varepsilon_1-\varepsilon_0}\left(\sin^2\tfrac{X}{2}+\sin^2\tfrac{Y}{2}\right)\right)\nonumber\\
    +&& \sigma^y\,\left(2\,\Theta_{1,0}^2\,\tfrac{U_0^2}{\varepsilon_1-\varepsilon_0}\,\sin\tfrac{X}{2}\,\sin\tfrac{Y}{2}\right)\,.
\end{eqnarray}
Landau-level mixing produces an overall downward energy shift and, more importantly, a $\sigma^y$ term that opens a gap of magnitude $\left|2\,\Theta_{1,0}^2\,\tfrac{U_0^2}{\varepsilon_1-\varepsilon_0}\right|$ at the former Dirac cones. Considering higher Landu levels gives corrections to the gap size but do not close it or change the Chern number.

We can now analytically classify the bands topologically. The two-band Hamiltonian gives Chern numbers $(-1/2,1/2)$, typical of a single Dirac cone \cite{Bernevig2006Dec}.T Adding the basis-state contribution $(1/2,1/2)$ yields total Chern numbers $(0,1)$, again in agreement with the numerical calculation.

Unlike the case for $\phi \leq 0$, the trivial band is now fully separated for an infinitesimal strength of the potential $U_0$. This is a satisfying result, that tells us that once the trivial band appears at $\phi=1$, it will persist for all values of the flux.

\section{\label{sec:tb}Tight-binding model}

Now that we have established the existence of a trivial band, given a sufficiently strong periodic potential, the question remains how similar or dissimilar this band is to the Hofstadter butterfly obtained from a tight-binding model with Peierls phases \cite{hofstadter_energy_1976}.

Our second key result of this paper is the surprising realization that the Hofstadter butterfly is \textit{not} realized in our realistic exact calculation. The Hofstadter butterfly fails in capturing two aspects. One is that, due to the large magnetic field, the size and shape of orbitals change; and therefore a tight-binding model constructed at zero field supplemented by Peierls phases gives an incorrect spectrum. This aspect will be discussed in Sec.~\ref{Sec:TightBindingParameters} when we analyse the tight-binding parameters. The second, more fundamental, difference between the exact butterfly and Hofstadter's is in the sequence of Chern numbers. In addition to the changed bandwidths and gap positions, also the \textit{topology} of the spectrum has changed. We find that this is due to a sign change in the diagonal hopping parameter, which is discussed in Sec.~\ref{sec:diagonalt}.

The full comparison between Hofstadter physics and the exact results requires a Wannierization of the electronic eigenstates. We will therefore focus on this procedure first. 

\subsection{Wannierization}
\label{Sec:Wannierization}

\begin{figure}
\centering
    \includegraphics[width=\linewidth]{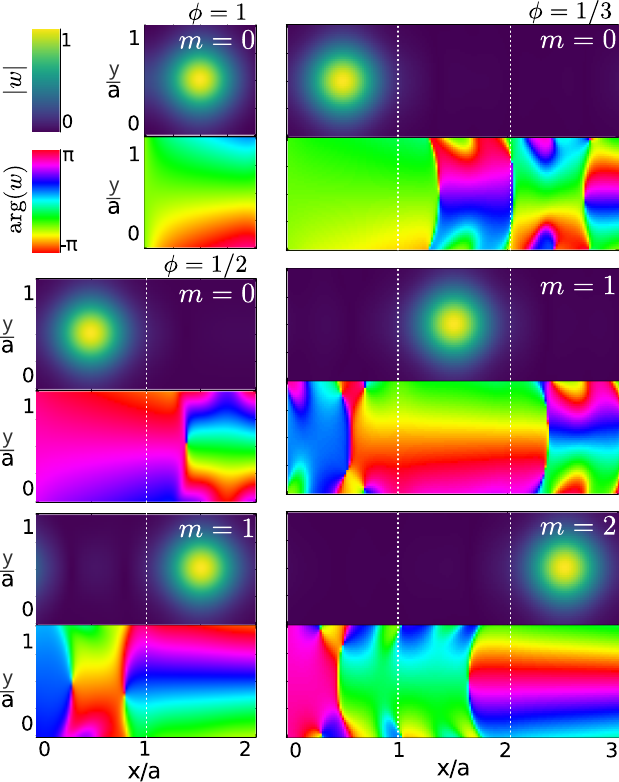}
    \caption{Wannier functions in the magnetic unit cell at $|U_0|=0.05\,\mathrm{eV}$ and $\phi=1$, $1/2$, and $1/3$. Each panel contains $q$ Wannier orbitals and shows their amplitude, normalized to unity, and phase.}
    \label{fig:wannier_functions}
\end{figure}

The electronic eigenstate densities are localized, approximately Gaussian functions that become narrower with increasing potential strength $U_0$ or magnetic flux. This motivates a Wannierization of the lowest Hofstadter-like band, which is topologically trivial as a whole despite its nontrivial subband structure. The outcome of a Wannierization procedure is to find a set of localized Wannier orbitals and the corresponding tight-binding parameters. In our case, we will find such set of orbitals and parameters \textit{for each separate flux} -- to be able to compare to the Hofstadter model which has \textit{the same} tight-binding parameters for all values of flux.

The greatest challenge in finding relevant Wannier orbitals is the gauge freedom inherent in Bloch wavefunction. We fix the gauge of the eigenstates by projecting them onto a set of trial wave functions, following Sec.~II.B of Ref.~\cite{marzari_maximally_2012}.

\begin{figure*}
    \includegraphics[width=\textwidth]{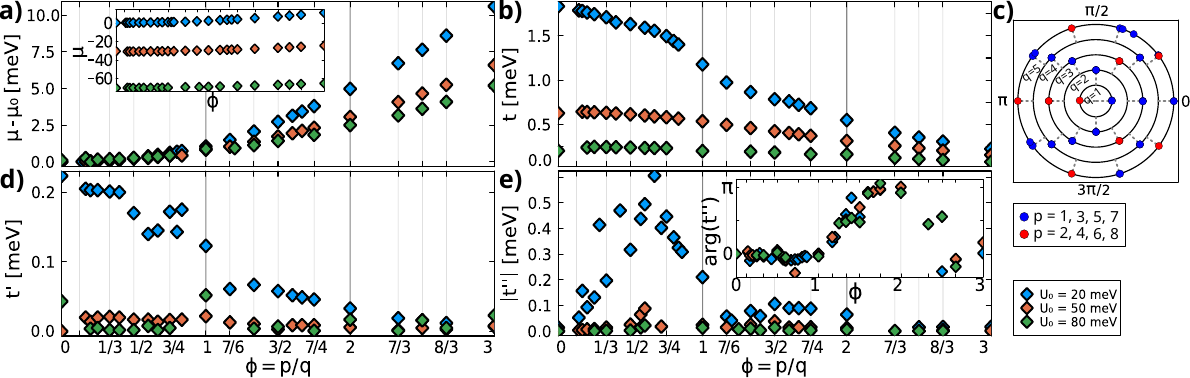}
    \caption{\label{fig:wannier_superfig} \textbf{(a)} Chemical potential relative to its zero-field value as a function of flux for three potential strengths: one near the Umklapp energy, one intermediate, and one deep in the Hofstadter regime. The inset shows the unsubtracted values. \textbf{(b)} Nearest-neighbor hopping amplitude as a function of flux. \textbf{(c)} Numerically obtained phases of the full nearest-neighbor hopping along $y$, plotted on unit circles for different $q$, in red and blue circles. The gray dashed lines show the phases expected for straight-line Hofstadter hopping between potential minima and they agree closely with the numerical values. \textbf{(d)} Axial next-nearest-neighbor hopping amplitude. \textbf{(e)} Absolute value of the complex diagonal next-nearest-neighbor hopping; the inset shows its complex phase.}
\end{figure*}

The question is which trial wavefunctions are best. A natural starting point would be to use Gaussian wave functions, especially at large $|U_0|$, because they are Fock-Darwin solutions $\ket{\psi_h}$ for a spinless electron in a harmonic potential and a magnetic field \cite{fock_bemerkung_1928}. Specifically, $\ket{\psi_h}\propto e^{-\bm{r}^2/(2\sigma^2)}\,e^{-i\pi\phi xy/a^2}$, where $\sigma^2=\tfrac{\hbar}{m_e}(\omega^2+\tfrac{1}{4}\omega_c^2)^{-1/2}$, $\omega_c$ is the cyclotron frequency, and $\omega(U_0)$ is the natural frequency obtained by expanding the periodic potential about a minimum. Note that this simple harmonic approximation already reveals that Wannier orbitals will become \textit{narrower} with increasing magnetic field.

In practice, however, we obtained more consistent results from using a projection on the zero-momentum ($X=Y=0$) Bloch wave functions in the lowest Hofstadter band. These states are periodically localized at sites of the magnetic unit cell. We truncate each Bloch function to the ordinary lattice unit cell in which it is localized and use the result as a trial function. The advantage of this approach is that they naturally inherit the correct real-space phase structure from the zero momentum Bloch functions. 

Therefore, we denote $\ket{g_m}$ as the trial wavefunctions obtained from the truncated Bloch wavefunctions at zero momentum. The overlap matrix between the normalized eigenstates and trial functions is $(A_{XY})_{m'm}=\braket{\psi_{m'XY}|g_m}$, and the overlap matrix in the projected basis is $(S_{XY})_{m'm}=(A_{XY}^\dagger A_{XY})_{m'm}$. The Wannier wave functions can then be expressed as Fourier transforms of the gauge-fixed Bloch wave functions:
\begin{align}
    \ket{w_{\bm{R},m}} = \tfrac{1}{4\pi^2} \int_0^{2
    \pi q}\int_0^{2\pi} &dXdY\; e^{i(XR_y+YR_x)} \\ & \times\left(A_{XY}S^{-\tfrac{1}{2}}_{XY}\right)_{mm'} \ket{\psi_{XY,m'}}. \nonumber
\end{align}
The resulting Wannier states for flux values $\phi=1$, $1/2$ and $1/3$ are shown in Fig.~\ref{fig:wannier_functions}. Numerically, we replace the integrals by finite sums over a real-space and momentum grids. The Wannier densities respect magnetic-unit-cell translation symmetry. 

\subsection{Tight-binding parameters}
\label{Sec:TightBindingParameters}

\begin{figure}
\centering
    \includegraphics[width=\linewidth]{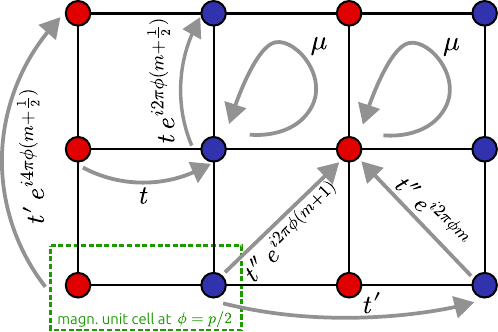}
    \caption{Illustration of the tight-binding model lattice and hopping parameters (here for $\phi=1/2$), as given in Eq.~\eqref{Eq:EffectiveTightBindingModel}.}
    \label{fig:reconstructed}
\end{figure}

\begin{figure*}
    \includegraphics[width=\textwidth]{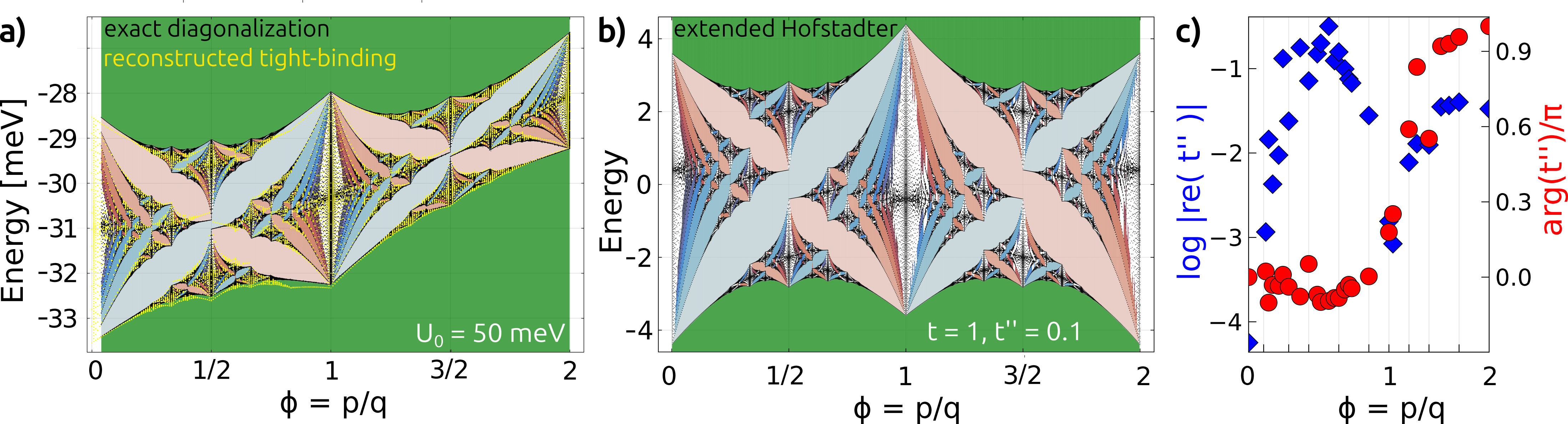}
    \caption{\label{fig:compare_butterflies} \textbf{(a)} The original spectrum at $U_0=50\,\mathrm{meV}$ from exact diagonalization in the numerical Landau level basis in black dots, and the reconstructed spectrum from the Wannierized extended tight-binding model in yellow dots. A small numerical error, arising from the truncation of the hopping range, gives a deviation from the original spectrum but otherwise all important features such as bandwidth, center energy, and gaps are the same. \textbf{(b)} Spectrum of an extended Hofstadter model with nearest-neighbor hopping along $x$ and $y$ with amplitude $t=1$, as well as diagonal next-nearest-neighbor hopping with amplitude $t''=0.1$. In both (a) and (b), the colors in the gaps correspond to Chern numbers, as described in Fig.~\ref{fig:spectra}. Notably, the two plots differ in Chern number in the central gap at $\phi=3/2$. \textbf{(c)} Log of the absolute value and argument of the diagonal next-nearest-neighbor hopping parameter }
\end{figure*}

Once the Wannier basis has been obtained at a flux $\phi=p/q$, the on-site energies and hopping amplitudes follow directly. The amplitude for hopping from orbital $m$ at $\bm{R}$ to orbital $m'$ at $\bm{R}'$ is
\begin{widetext}
\begin{eqnarray}
    t^{mm'}_{\bm{R},\bm{R}'} = \bra{w_{m'\bm{R}'}} H \ket{w_{m\bm{R}}} = \int_0^{2\pi}dY\int_0^{2\pi q}dX\,e^{-i\left(X(R_y-R_y') + Y(R_x-R_x')\right)} \left(S^{-\tfrac{1}{2} \dagger}_{XY} A_{XY}^\dagger \,\text{diag}(E_j) \, A_{XY}S^{-\tfrac{1}{2}}_{XY}\right)_{mm'}.
\end{eqnarray}
\end{widetext}
We calculate the tight-binding parameters of the lowest Hofstadter-like band for $p,q\in\{1,\ldots,8\}$ with $0<p/q<3$, at three potential strengths above the critical value where the lowest band is fully separated from the rest of the spectrum. The resulting hopping parameters are summarized in Fig.~\ref{fig:wannier_superfig}.

The on-site energy $\mu$, shown in Fig.~\ref{fig:wannier_superfig}(a), decreases with increasing potential strength because the potential minima become deeper. By contrast, increasing the magnetic flux raises the on-site energy. Within the harmonic approximation about a minimum, the field increases the effective oscillator frequency and hence the ground-state energy, which scales as $\sqrt{\alpha U_0+\beta\phi^2}$.

The amplitude $|t|$ of the nearest-neighbor hopping decreases with increasing potential strength and magnetic flux because both enhance orbital confinement, which can be clearly seen in Fig.~\ref{fig:wannier_superfig}(b). The phases of hoppings along $x$ can appear arbitrary because numerical Wannierization may assign a phase $\chi_m$ to each orbital with index $m$. For $q>1$, nearest-neighbor hopping along $x$ connects different Wannier orbitals and therefore acquires an additional gauge-dependent phase $\chi_m-\chi_{m\pm1}$, which does not affect the spectrum. Hopping along $y$, however, connects orbitals with the same internal index, so this ambiguity cancels and the resulting phase follows a regular, physical pattern. We choose the vector-potential gauge to vanish at the origin, while the potential minima lie at half-integer multiples of $a$. Straight-line hopping between minima of nearest-neighbors along $y$ is therefore expected to acquire the Hofstadter phase $\theta_m^H(\phi)=2\pi\phi(m+\tfrac{1}{2})$.  Within numerical accuracy, this is precisely the phase we observe as shown in Fig.~\ref{fig:wannier_superfig}(c) -- a notable result because no hopping phases were assumed in the continuum Hamiltonian of Eq.~\eqref{eq:hamiltonian_continuum}. This provides a beautiful verification of the Peierls minimal coupling argument for a tight-binding model.

Axial next-nearest-neighbor hopping -- the next nearest-neighbor along the $x$ or $y$ axes of the lattice, shown in Fig.~\ref{fig:wannier_superfig}(d) -- behaves qualitatively like nearest-neighbor hopping. Its amplitude is roughly an order of magnitude smaller, and is therefore more susceptible to numerical noise. However, the complex phase of the axial hopping parameter is fully consistent with the Peierls description. 

Diagonal next-nearest-neighbor hopping differs from the straight-line expectation. We focus on hopping along the positive diagonal, $\Delta x=\Delta y=a$. Because this hopping connects different Wannier orbitals, we remove the numerical gauge phase by subtracting the corresponding nearest-neighbor hopping phase along $x$. Straight-line hopping in an extended Hofstadter model would have the phase $\theta_m^{H\text{-}d}(\phi)=2\pi\phi(m+1)$. For $0<\phi<1$, where the band contains contributions from several Landau levels, the numerical phase closely follows this expression. For $\phi>1$, where the band is instead formed from subbands of a single Landau level, the residual phase $\arg(t'')=\arg(\bra{w_{m+1,(x_0+a,y_0+a)}}H\ket{w_{m,(x_0,y_0)}})-\theta_m^{H\text{-}d}$ oscillates, producing a smooth sign reversal in the complex plane between integer fluxes, as shown in Fig.~\ref{fig:wannier_superfig}(e).

As a result, we can reconstruct our full spectrum with an extended tight-binding Hofstadter model, whose Hamiltonian is given by 
\begin{align}
    H_{TB} &= \tfrac{\mu}{2}\sum_j c^\dagger_{m,j} c_{m,j} - t\sum_{\langle ij\rangle_x} c^\dagger_{m+1,i} c_{m,j} \nonumber \\
    &- t\sum_{\langle ij\rangle_y} e^{i2\pi\phi(m+\tfrac{1}{2})} c^\dagger_{m,i} c_{m,j} - t'\sum_{\langle\langle ij\rangle\rangle_x} c^\dagger_{m+2,i} c_{m,j} \nonumber \\
    &- t'\sum_{\langle\langle ij\rangle\rangle_y} e^{i4\pi\phi(m+\tfrac{1}{2})} c^\dagger_{m,i} c_{m,j} \nonumber \\
    &- t''\sum_{\langle\langle ij\rangle\rangle_{\nearrow}} e^{i2\pi\phi(m+1)} c^\dagger_{m+1,i} c_{m,j} \nonumber \\
    &- t''\sum_{\langle\langle ij\rangle\rangle_\nwarrow} e^{i2\pi\phi m} c^\dagger_{m-1,i} c_{m,j} + \text{h.c.}
    \label{Eq:EffectiveTightBindingModel}
\end{align}
where $\mu,t,t'$, and $t''$ depend on $\phi$ and are obtained by linearly interpolating the data in Fig.~\ref{fig:wannier_superfig}. While $t$ and $t'$ are real, only $t''$ is complex in our reconstruction of the spectrum shown in Fig.~\ref{fig:compare_butterflies}(a). 

The tight-binding model Eq.~\eqref{Eq:EffectiveTightBindingModel} is the third key result of our paper. With this, we show that a tight-binding model can still be realized in strong magnetic fields provided the periodic potential is sufficiently strong. Such a tight-binding model can serve as a realistic start for many-body calculations that would have been otherwise challening in a large magnetic field.

\subsection{\label{sec:diagonalt}Diagonal hopping anomaly}

In addition to its strongly flux-dependent magnitude, we found that the diagonal hopping $t''$ exhibits a complex phase with nontrivial flux dependence. This phase has direct topological consequences. At fluxes $\phi=p/2$ with $p$ odd, the nearest-neighbor Hofstadter model has $\pi$-flux Dirac points. Diagonal hopping generates a Dirac mass and opens a gap, analogous to the gap produced by Landau-level mixing at $\phi=2$ in Fig.~\ref{fig:SW_gap}. The sign of this mass determines the Chern number of the gap. By contrast, axial next-nearest-neighbor hopping produces an identity-like, momentum-dependent dispersion that breaks the $E\leftrightarrow-E$ symmetry but does not enter the Dirac mass.

In the conventional extended Hofstadter model, $t''$ is real and flux independent, with the magnetic-field dependence contained entirely in the Peierls phase. A diagonal bond forms a triangular loop enclosing half an elementary plaquette, so increasing the flux by one quantum changes the phase around this loop by $\pi$. The diagonal contribution to the Dirac mass therefore reverses sign under $\phi\mapsto\phi+1$, and the model is periodic only under $\phi\mapsto\phi+2$. Consequently, it has a $C=1$ gap at $\phi=1/2$ but a $C=-1$ gap at $\phi=3/2$, as shown in Fig.~\ref{fig:compare_butterflies}(b).

This differs from the exact continuum spectrum, in which both gaps carry $C=1$. The result at $\phi=3/2$ follows directly from the inverse-butterfly structure of the broadened Landau level. Its three subbands have Chern numbers $(+1,-1,+1)$ and therefore produce the cumulative gap sequence $(1,0,1)$ \cite{thouless_quantized_1982}. This sequence remains unchanged as the potential is increased, provided that the corresponding gaps do not close, and is distinct from the $(-1,0)$ sequence of the extended Hofstadter model.

The same distinction appears at $\phi=1/4$ and $\phi=5/4$. The nearest-neighbor model has a central Dirac touching at both fluxes, but a fixed diagonal hopping generates masses of opposite sign, whereas the exact calculation gives the same Chern-gap sequence. More generally, this is a feature of fluxes $\phi=p/q$ with even $q$ and odd $p$. The nearest-neighbor Hofstadter model has a central Dirac touching whose opening is controlled by longer-range hopping. At odd $q$, all there are no band touchings even in the nearest-neighbor Hofstadter model and therefore a small diagonal hopping might alter the gap sizes but not the Chern number. We observe correspondingly that $\phi\to\phi+1$ always preserves the gap Chern number sequence for odd $q$.

The argument can be extended to longer-range hopping. Consider a straight-line hopping with displacement $\Delta\bm{r}=(\Delta x,\Delta y)$. Together with an axis-aligned nearest-neighbor path between the same sites, it encloses an area $\Delta x \Delta y/2$. Under $\phi\mapsto\phi+1$, its phase relative to the nearest-neighbor path therefore changes by
\begin{align}
    e^{i\pi \Delta x \Delta y}=(-1)^{\Delta x \Delta y}.
\end{align}

In principle, the full hopping amplitude is a superposition of all possible paths, not just the straight-line. Interference and competition between different hopping paths may account for the smooth complex phase found numerically, but the compensating sign change itself is required by the topology of the exact spectrum.

\section{Conclusion}
We have followed the evolution of continuum Landau levels in a square periodic potential from weak Landau level broadening to an isolated Hofstadter-like band. This transition is controlled by gap closings at $\phi=1/q$, such that the appearance of a trivial band happens above a critical potential strength $U^c$ at flux $\phi =1$. Once the lowest band is separated from the rest of the spectrum, it allows for a Wannier description whose effective tight-binding parameters depend strongly on magnetic flux. In particular, longer-range hopping need not follow a single straight Peierls path: the diagonal hopping can change sign, indicating that interference among several real-space paths becomes important.

Our Wannierization provides a useful first effective model, but its numerical accuracy could be improved by constructing maximally localized orbitals independently at each flux. Such a construction would test the complex diagonal-hopping phase more stringently and allow for a more reliable treatment of many-body interactions. Because the magnetic field controls both the bandwidth and the hopping hierarchy, the present mechanism may provide a route to field-tuned correlation effects, for example a magnetic field tuned Mott transition caused by the field-induced bandwidth change. 

The critical value of the potential for the emergence of a trivial band $U^c$ is very close to the Umklapp energy scale $E_U$, though we could not extract an analytical relation between these two. We expect that the precise relation is dependent on the lattice geometries. This naturally brings us to different lattice geometries beyond square - such as rectangular or hexagonal. Naturally, the procedure is the same as presented in our work here, only the precise commensurability conditions might differ.

The electronic spectrum for hexagonal lattices are naturally relevant for moiré systems, such as twisted bilayer graphene or twisted transition-metal dichalcogenides, who are in vast majority hexagonal  \cite{Das2022May}. Field-, pressure- or twist angle-tuning allows for an effective change in the periodic potential strength, and as such allows experimental control over the eclosion of Landau caterpillars into butterflies \cite{Rohringer2025Nov,Yankowitz2019Jan}.

\begin{acknowledgments}
We thank Christophe Mora, and Oskar Vafek for stimulating discussions. We thank Dario Rossi for coming up with the term `caterpillar'. 
This research was funded by the SNSF via Starting Grant TMSGI2~211296.

\end{acknowledgments}

\newpage
\appendix

\begin{widetext}
\section{Calculation of matrix elements} \label{ap:matrix_el}
In this appendix, we show how we arrive at the functions
\begin{eqnarray}
    \Theta_{n,n'}(\phi) = e^{-\frac{\pi}{2\phi}}\sqrt{\frac{\min(n,n')!}{\max(n,n')!}} \left(\frac{\pi}{\phi}\right)^{\frac{|n'-n|} {2}} L^{|n'-n|}_{\min(n,n')}\left(\tfrac{\pi}{\phi}\right)\,,
\end{eqnarray}
which appear in the Hamiltonian in Eq.~\eqref{eq:hamiltonian_1d}. As stated in the main text, $L^{(\alpha)}_n(x)=\frac{x^{-\alpha}e^x}{n!}\frac{\partial^n}{\partial x^n}\left(e^{-x}x^{n+\alpha}\right)=\sum_{k=0}^n\tbinom{n+\alpha}{\alpha+k}\tfrac{(-x)^k}{k!}$ are the generalized Laguerre polynomials. Because the potential-free Hamiltonian is already diagonal in the Landau-level basis, we need only calculate the matrix elements of the cosine potential. We begin with its $x$-dependent component and decompose it into exponentials. In terms of the orthonormal Hermite functions $\eta_n(z)=(2^n n! \sqrt{\pi})^{-1/2}e^{-z^2/2}H_n(z)$, the Landau-level wave functions are
\begin{eqnarray}
    \ket{LL_{n,k_y}(x,y)}=\tfrac{1}{\sqrt{l_B\,a}}\eta_n(\xi)\,e^{ik_yy}
\end{eqnarray}
where $\xi = \tfrac{x}{l_B}-k_yl_B$. Therefore,
\begin{eqnarray}
    \bra{LL_{n',k_y'}(x,y)}&&\,e^{i2\pi x/a}\,\ket{LL_{n,k_y}(x,y)} =   \int_{0}^a\frac{dy}{a} \int_{-\infty}^\infty\frac{dx}{l_B} \,\eta_{n'}(\xi')\,\eta_n(\xi)\,e^{i2\pi x/a}\,e^{iy(k_y-k_y')} \\
    =&& \delta_{k_y,k_y'} \,(2^{n+n'} n!n'! \pi)^{-1/2} e^{i\xi_0k_yl_B}e^{-\xi_0^2/4}\,\int_{-\infty}^\infty  d\xi \,H_{n'}(\xi)\,H_n(\xi)\,e^{-(\xi-i\tfrac{\xi_0}{2})^2}  \nonumber\\
    =&& \delta_{k_y,k_y'} \,(2^{n+n'} n!n'! \pi)^{-1/2} e^{i\xi_0k_yl_B}e^{-\xi_0^2/4}\,\sum_{k=0}^{\min(n,n')}\tbinom{n'}{k}\tbinom{n}{k} 2^k k! \int_{-\infty}^\infty  d\xi \,H_{n+n'-2k}(\xi)\,e^{-(\xi-i\tfrac{\xi_0}{2})^2}  \nonumber\\
    =&& \delta_{k_y,k_y'} \,\sqrt{\tfrac{2^{n+n'}}{n!n'!}} e^{i\xi_0k_yl_B}e^{-\xi_0^2/4}\,\sum_{k=0}^{\min(n,n')}\tbinom{n'}{k}\tbinom{n}{k} 2^{-k} k!  \left(\tfrac{i\xi_0}{2}\right)^{n+n'-2k} \nonumber
\end{eqnarray}
where $\xi_0=2\pi l_B/a=\sqrt{2\pi/\phi}$ and we have used the Hermite polynomial identities
\begin{eqnarray}
    H_n(z)H_m(z)=\sum_{k=0}^{\min(n,m)}2^kk! H_{n+m-2k}(z)\,;\\
    \int_{-\infty}^\infty  dt \,H_n(t)\,e^{-(t-z)^2}=2^n\sqrt{\pi}\,z^n\,.
\end{eqnarray}
Hence,
\begin{eqnarray}
    \bra{LL_{n',k_y'}(x,y)}&&\cos(2\pi x/a)\,\ket{LL_{n,k_y}(x,y)} = \\
    =&&\tfrac{1}{2}\,\delta_{k_y,k_y'} \,\sqrt{\tfrac{2^{n+n'}}{n!n'!}} e^{-\xi_0^2/4}\,\sum_{k=0}^{\min(n,n')}\tbinom{n'}{k}\tbinom{n}{k} 2^{-k} k!\, \left[ e^{i\xi_0k_yl_B}\,\left(\tfrac{i\xi_0}{2}\right)^{n+n'-2k} + \text{c.c.}\right]\nonumber\\
    =&&\,\delta_{k_y,k_y'} \,\cos(\xi_0k_yl_B+\pi\tfrac{n+n'}{2})\, e^{-\xi_0^2/4} \,\sqrt{\tfrac{2^{n+n'}}{n!n'!}}(\tfrac{\xi_0}{2})^{n+n'}\,\sum_{k=0}^{\min(n,n')}\tbinom{n'}{k}\tbinom{n}{k}\, k!\,\left(\tfrac{-\xi_0^2}{2}\right)^{-k}\nonumber\\
    =&&\,\delta_{k_y,k_y'} \,\cos(2\pi k_yl_B^2/a+\pi\tfrac{n+n'}{2})\, e^{-\tfrac{\pi}{2\phi}} \,\sqrt{\tfrac{1}{n!n'!}}\left(\sqrt{\tfrac{\pi}{\phi}}\right)^{|n'-n|}\,(-1)^{\min(n,n')}\min(n,n')!\,L^{|n'-n|}_{\min(n,n')}(\tfrac{\pi}{\phi})\nonumber\\
    =&&\,\delta_{k_y,k_y'} \,\cos(2\pi k_yl_B^2/a+\pi\tfrac{|n'-n|}{2})\,\Theta_{n,n'}(\phi)\nonumber\,.
\end{eqnarray}
Recall that
\begin{align}
    2\pi l_B^2/a = \frac{a}{\phi}
\end{align}
so that the $x$-component of the potential is periodic in $k_y$ with period $\frac{2\pi}{a} \phi$.

For the $y$-component of the potential, 
\begin{eqnarray}
    \bra{LL_{n',k_y'}(x,y)}&&\,e^{i2\pi y/a}\,\ket{LL_{n,k_y}(x,y)} =  \int_{0}^a\frac{dy}{a} \int_{-\infty}^\infty\frac{dx}{l_B} \,\eta_{n'}(\xi')\,\eta_n(\xi)\,e^{iy(k_y-k_y'+\tfrac{2\pi}{a})} \\
    =&& \delta_{k_y',k_y+\tfrac{2\pi}{a}}  \int_{-\infty}^\infty  d\xi \,\eta_{n'}(\xi-\xi_0)\,\eta_n(\xi) \nonumber\\
    =&& \delta_{k_y',k_y+\tfrac{2\pi}{a}} \,\times\Bigg\{\begin{array}{ll}
         \sqrt{n!n'!}\,e^{-\xi_0^2/4}\, (\tfrac{-\xi_0}{\sqrt{2}})^{n'-n}\sum_{k=0}^n \tfrac{(-\xi_0^2/2)^{-k}}{k!(n-k)!(k-n+n')}&\quad\text{if}\quad n'\geq n  \\
         \sqrt{n!n'!}\,e^{-\xi_0^2/4}\, (\tfrac{\xi_0}{\sqrt{2}})^{n-n'}\sum_{k=0}^{n'} \tfrac{(-\xi_0^2/2)^{-k}}{k!(n'-k)!(k-n'+n)}&\quad\text{if}\quad n'< n 
    \end{array} \nonumber\\
    =&& \delta_{k_y',k_y+\tfrac{2\pi}{a}} e^{-\tfrac{\pi}{2\phi}}\times\Bigg\{\begin{array}{ll}
         \sqrt{\tfrac{n!}{n'!}}\, \left(-\sqrt{\tfrac{\pi}{\phi}}\right)^{n'-n}L_n^{n'-n}(\tfrac{\pi}{\phi})&\quad\text{if}\quad n'\geq n  \\
         \sqrt{\tfrac{n'!}{n!}}\, \left(\sqrt{\tfrac{\pi}{\phi}}\right)^{n-n'}L_{n'}^{n-n'}(\tfrac{\pi}{\phi})&\quad\text{if}\quad n'< n 
    \end{array} \nonumber\\
    =&& \delta_{k_y',k_y+\tfrac{2\pi}{a}}\, (-1)^{\tfrac{n'-n+|n'-n|}{2}} \Theta_{n,n'}\nonumber
\end{eqnarray}
where we have used the Hermite function identity 
\begin{equation}
    \int_{-\infty}^\infty dz\, \eta_m(z-z_0)\eta_n(z) = \sqrt{n!m!}\,(\tfrac{-z_0}{\sqrt{2}})^{m-n}\sum_{k=0}^{n}\tfrac{(-z_0^2/2)^k}{k!(n-k)!(k-n+m)!}\quad\text{for} \quad m\geq n\,.
\end{equation}
Now for the cosine, we get
\begin{eqnarray}
    \bra{LL_{n',k_y'}(x,y)}&&\cos(2\pi y/a)\,\ket{LL_{n,k_y}(x,y)} = \\
    =&&\tfrac{\Theta_{n,n'}}{2}\left(\delta_{k_y',k_y+\tfrac{2\pi}{a}}(-1)^{\tfrac{n'-n+|n'-n|}{2}} + \delta_{k_y',k_y-\tfrac{2\pi}{a}}(-1)^{\tfrac{n-n'+|n'-n|}{2}}\right) \nonumber \\
    = && (-1)^{\tfrac{n'-n+|n'-n|}{2}} \tfrac{\Theta_{n,n'}}{2}\left(\delta_{k_y',k_y+\tfrac{2\pi}{a}}+ \delta_{k_y',k_y-\tfrac{2\pi}{a}}(-1)^{n-n'}\right)
\end{eqnarray}

\section{Chern numbers of the Landau level basis}\label{ap:chern_landau}
We write the basis states from Eq.~\eqref{eq:landau_basis}, omitting the $X$ and $Y$ indices and using the notation of Appendix~\ref{ap:matrix_el}, as
\begin{eqnarray}
    \ket{\psi_{nm}}=\tfrac{1}{2\pi\sqrt{l_B\,a}}\tfrac{1}{\sqrt{2K_y^{\max}+1}}\sum_{K_y=-K_y^{\max}}^{K_y^{\max}}  e^{-iK_yY} \, e^{iy(\tfrac{X}{qa}+\tfrac{2\pi p}{a}K_y+\tfrac{2\pi}{a}m)} \, \eta_n\left(\tfrac{x}{l_B} - l_B[\tfrac{X}{qa}+\tfrac{2\pi p}{a}K_y+\tfrac{2\pi}{a}m]\right)\,.
\end{eqnarray}
Therefore, we have for the Berry connection components
\begin{eqnarray}
    A^Y_{nm}=&&i\bra{\psi_{nm}}\partial_Y\ket{\psi_{nm}} \\
    =&&\tfrac{1}{2K_y^{\max}+1}\sum_{K_y,K_y'=-K_y^{\max}}^{K_y^{\max}} K_y\;  e^{-iY(K_y-K_y')} \,\int_0^a \frac{dy}{a}\, e^{iy\tfrac{2\pi p}{a}(K_y-K_y')} \int_{-\infty}^\infty d\xi \, \eta_n(\xi)\, \eta_n\left(\xi+\tfrac{2\pi p}{a}(K_y-K_y')\right) \nonumber \\
    =&&\tfrac{1}{2K_y^{\max}+1}\sum_{K_y,K_y'=-K_y^{\max}}^{K_y^{\max}} K_y\;  e^{-iY(K_y-K_y')} \,\delta_{K_y,K_y'} \,L_n\left(\tfrac{2\pi p}{a}(K_y-K_y')\right) \nonumber \\
    =&&\tfrac{1}{2K_y^{\max}+1}\sum_{K_y=-K_y^{\max}}^{K_y^{\max}} K_y \nonumber \\
    =&& 0\nonumber \,.
\end{eqnarray}
and
\begin{eqnarray}
    A^X_{nm}=&&i\bra{\psi_{nm}}\partial_X\ket{\psi_{nm}} = \tfrac{-1}{qa}\bra{\psi_{nm}}y\ket{\psi_{nm}} + \kappa_1\langle\psi_{nm}|\psi_{n+1,m}\rangle + \kappa_2\langle\psi_{nm}|\psi_{n-1,m}\rangle \\
    =&& \tfrac{-1}{qa}\bra{\psi_{nm}}y\ket{\psi_{nm}} \nonumber
\end{eqnarray}
where we used the basis orthogonality and the Hermite function identity 
\begin{equation}
    \partial_z \eta_n(z)= \sqrt{\tfrac{n}{2}}\,\eta_{n-1}(z) - \sqrt{\tfrac{n+1}{2}}\,\eta_{n+1}(z)
\end{equation}
such that $\kappa_{1,2}\in\mathbb{C}$ are some constants. Hence,
\begin{eqnarray}
    A^X_{nm} &&= -\tfrac{1}{qa}\tfrac{1}{2K_y^{\max}+1}\sum_{K_y,K_y'=-K_y^{\max}}^{K_y^{\max}}   e^{-iY(K_y-K_y')} \,\int_0^a \frac{dy}{a}\,y\, e^{iy\tfrac{2\pi p}{a}(K_y-K_y')} \int_{-\infty}^\infty d\xi \, \eta_n(\xi)\, \eta_n\left(\xi+\tfrac{2\pi p}{a}(K_y-K_y')\right) \nonumber \\
    &&=\tfrac{i}{2\pi pq}\tfrac{1}{2K_y^{\max}+1}\sum_{K_y,K_y'=-K_y^{\max}}^{K_y^{\max}}   e^{-iY(K_y-K_y')} \,\partial_{K_y}\left(\delta_{K_y,K_y'}\right) \,L_n\left(\tfrac{2\pi p}{a}(K_y-K_y')\right) \nonumber \\
    &&=\tfrac{-i}{2\pi pq}\tfrac{1}{2K_y^{\max}+1}\sum_{K_y,K_y'=-K_y^{\max}}^{K_y^{\max}}  \delta_{K_y,K_y'}\, \partial_{K_y}\left(e^{-iY(K_y-K_y')}\,L_n\left(\tfrac{2\pi p}{a}(K_y-K_y')\right)\right) + \text{const}\nonumber \\
    &&=\tfrac{-Y}{2\pi pq} + \text{const}\nonumber \,.
\end{eqnarray}
Integrating the Berry connection along the boundary of the parameter space leaves a single nonzero term, which gives a Chern number $C=1/p$ for every band of every Landau level. The Berry curvature is also uniform. Taking the limit $K_y^{\max}\to\infty$ does not change the result.

\end{widetext}

\section{Numerical}
\label{Appendix:Numerical}

Our numerical code is available on GitHub.\cite{github_landau_butterflies}

\bibliography{refs_fixed}

\end{document}